\documentclass[preprint,12pt]{elsarticle}

\usepackage{amssymb}
\usepackage{amsmath}

\newcommand{\approxpm}{\mathrel{\vcenter{
  \offinterlineskip\halign{\hfil$##$\cr
     \pm\cr\noalign{\kern2pt}\sim\cr\noalign{\kern-2pt}}}}}

\journal{Ultramicroscopy}

\begin{document}

\begin{frontmatter}





\title{Image registration of low signal-to-noise cryo-STEM data}

\author[CornellPhysics]{Benjamin H. Savitzky\corref{cor}}
\author[CornellPhysics]{Ismail El Baggari}
\author[CornellPhysics]{Colin Clement}
\author[CornellAEP]{Emily Waite}
\author[JHU]{John P. Sheckelton}
\author[JHU]{Christopher Pasco}
\author[Rutgers]{Alemayehu S. Admasu}
\author[Rutgers]{Jaewook Kim}
\author[Rutgers]{Sang-Wook Cheong}
\author[JHU]{Tyrel M. McQueen}
\author[CornellAEP,MichiganMatSci]{Robert Hovden}
\author[CornellAEP,Kavli]{Lena F. Kourkoutis\corref{cor}}

\address[CornellPhysics]{Department of Physics, Cornell University, Ithaca, NY 14853}
\address[CornellAEP]{School of Applied and Engineering Physics, Cornell University, Ithaca, NY 14853}
\address[JHU]{Department of Chemistry, The Johns Hopkins University, Baltimore, MD 21218}
\address[Rutgers]{Rutgers Center for Emergent Materials and Department of Physics and Astronomy, Rutgers University, Piscataway, NJ 08854, USA}
\address[Kavli]{Kavli Institute at Cornell for Nanoscale Science, Cornell University, Ithaca, NY 14853}
\address[MichiganMatSci]{Present Address: Department of Materials Science and Engineering, University of Michigan, Ann Arbor, MI 48109}
\cortext[cor]{Corresponding Author}

\begin{abstract}

Combining multiple fast image acquisitions to mitigate scan noise and drift artifacts has proven essential for picometer precision, quantitative analysis of atomic resolution scanning transmission electron microscopy (STEM) data.
For very low signal-to-noise ratio (SNR) image stacks -- frequently required for undistorted imaging at liquid nitrogen temperatures -- image registration is particularly delicate, and standard approaches may either fail, or produce subtly specious reconstructed lattice images.
We present an approach which effectively registers and averages image stacks which are challenging due to their low-SNR and propensity for unit cell misalignments.
Registering all possible image pairs in a multi-image stack leads to significant information surplus.
In combination with a simple physical picture of stage drift, this enables identification of incorrect image registrations, and determination of the optimal image shifts from the complete set of relative shifts.
We demonstrate the effectiveness of our approach on experimental, cryogenic STEM datasets, highlighting subtle artifacts endemic to low-SNR lattice images and how they can be avoided.
High-SNR average images with information transfer out to 0.72~\AA~are achieved at 300 kV and with the sample cooled to near liquid nitrogen temperature.

\end{abstract}

\begin{keyword}

Scanning Transmission Electron Microscopy (STEM) \sep
Cryogenic STEM (cryo-STEM) \sep
Rigid registration \sep
Image reconstruction \sep
Atomic Resolution \sep
Low signal-to-noise ratio (SNR)




\end{keyword}

\end{frontmatter}


\section{Introduction}
\label{S:intro}

Imaging atomic structures with sub-angstrom resolution and sub-picometer precision is now possible in modern scanning transmission electron microscopes (STEMs).
While advances in aberration correction have enabled sub-angstrom electron probes \cite{haider2000upper,Batson2002,tanaka2008present}, making full use of these narrow electron beams has required optimizing the stability of the microscope, sample stage, and room environment \cite{Muller2006}.
To minimize the effect of any remaining mechanical, electromagnetic, thermal, and acoustic instabilities and to improve the signal-to-noise ratio (SNR) of the final image, a variety of post-processing algorithms have been developed, and have proven essential for high precision, quantitative STEM analysis \cite{Sang2014,Berkels2014,Ophus2016,ning2017}.

STEM imaging of samples cooled to liquid nitrogen temperatures (cryo-STEM) opens the possibility of characterizing the atomic structure of electronic materials across phase transitions, probing processes at solid-liquid interfaces, examining the structure of cells and other biological systems across a wide range of sample thicknesses, or controlling carbon contamination effects \cite{elbaggari2017,zachman2016,wolf2014cryo,spoth2017dose,elad2017detection}.  
Currently, cooling a sample while preserving the ability to align along a crystallographic axis is only possible with side entry cryo holders, in which the sample stage is in thermal contact with a liquid nitrogen bath, resulting in increased stage drift and additional noise due to cryogen bubbling.
Bubbling can be minimized by ensuring good thermal isolation between the cryogen and the environment, and maintaining a clean dewar to prevent bubble nucleation.
Drift can be minimized by allowing sufficient time for the stage to settle, however, is difficult to fully eliminate.
The effect of sample drift can be mitigated by acquiring many images with very short frame times, and subsequently registering and averaging the resultant stack of images \cite{kimoto2010}.
However, the frame times required (often $<1$ s) can yield very low-SNR data, complicating image registration.
Particularly challenging datasets, such as nearly perfectly translationally symmetric images (e.g. featureless lattices), can exacerbate the problem by inducing unit cell misalignments between image pairs.
This precise situation arises in many solid state systems where picometer-precision atomic position fitting is most relevant for probing the underlying physics, much of which only emerges in low temperature phases, including multiferroics, charge density wave systems, and high temperature superconductors \cite{Eerenstein2006,Gruner1988,tinkham1996introduction}.

Here, we present an image registration approach that is optimized for difficult, low-SNR cryo-STEM images, which often cannot be registered successfully by other means.
We introduce an approach which does not rely on a single reference image, but instead uses all possible combinations of image correlations to determine the optimal shifts.
Incorrect correlations, which plague low-SNR data, are then identified and handled by enforcing physical consistency from the surplus of information present in registrations of all image pairs.
Our approach accounts for sampling errors which can result in unit-cell jumps in translationally symmetric data, minimizing the possibility of these artifacts.
As difficult datasets often involve exploring multiple combinations of registration parameters, our implementation is designed to be both fast and flexible, allowing straightforward variation of real space boundary condition handling, Fourier space masking, choice of correlation function (cross correlation, mutual correlation, phase correlation \cite{VanHeel1992,Foroosh2002}), correlation maximum determination, and outlier removal methods.
The implementation outputs a brief report on each registration performed which facilitates quick determination of success or failure, both qualitatively and quantitatively.

The source code is available as a free, open source Python package with a modular, extensible structure, designed for either interactive use through the Jupyter notebook, for automated batch processing, or with a freely available graphical user interface.
Code can be obtained through the Python Package Index via XXXXX, and all source code is freely available on github at XXXXX \footnote{Please note that links to code will be available with the final publication.}.

\subsection{Approaches to image correction}

Rapid progress in aberration corrected STEM in the early and mid 2000’s was followed by various approaches and implementations to correcting image artifacts or distortions.  
The earliest approaches involved deconvolution of the probe and object functions \cite{Watanabe2002,Nakanishi2002}.
STEM and TEM images of identical regions were used to correct for non-orthogonal or continuously warped regions in the STEM data, in either reciprocal or real space \cite{Nakanishi2002,Recnik2005}.
Others determined and corrected for systematic distortions in their particular microscopes by examining the similarities in strain fields across many lattice images of many sample regions using geometric phase analysis \cite{Sanchez2006,Hytch1998}.

Scan noise, offsets in the starting position of each scan line, is particularly difficult to diagnose and correct.
Scan noise results in blurring of the Bragg peaks in fast Fourier transforms (FFTs) of lattice images along the slow scan direction, thus one approach to scan noise correction involves analyzing the phase information in these streaks to directly extract and correct for scanning offsets \cite{Braidy2012}.
Alternative approaches include shifting pixels along the fast scan direction to maximize their cross correlation with a section of pixels in the adjacent rows, and rearranging rows of pixels vertically to ensure the intensity of each atomic column decreases monotonically from its center \cite{Jones2013}.

Methods to align, or register, images span electron microscopy, scanned probe miscoscopies, medical imaging, cartography, computer vision, and many other fields \cite{Zitova2003,modersitzki2004}.
The fundamental limits of the general image registration problem have been been explored at low- and high-SNRs for single and multiple image registrations \cite{Robinson2004,Aguerrebere2016,vural2013}.
Efficient, high fidelity registration is required for cryo-TEM \cite{Joyeux2002,Shatsky2009,Rubinstein2015}.
In STEM, image registration and averaging tends to average out both scan noise and Poisson noise, and several approaches have been developed.
Rotation of the scan direction has been used to diagnose and correct for constant or linearly varying sample drift \cite{Sang2014}.
Registration methods which allow for continuous, or `non-rigid', distortion of the probe position during scanning have been developed and applied to obtain sub-pm precision identification of atomic positions \cite{Berkels2014,Yankovich2014}.
Another rotating scan approach determines and corrects for shifts in the initial position of each scan line by leveraging the superior information transfer along the fast scan direction, comparing local information in scans rotated by 90 degrees, and ultimately weighting information in Fourier space more heavily along the fast scan direction of each image before averaging \cite{Ophus2016}.

The approach here is comparatively simple.
We begin with the assumption that all images in an acquisition series are identical, save for a translational offset due to drift of the sample stage.
While this ignores the complicated and real effects of continuous image distortions from scanning offsets, or higher frequency stage position variations, we find that this simpler approach is well suited to low-SNR cryo-STEM imaging, in which particular care is required to avoid subtle artifacts from incorrect registration.
Here, we document such subtle artifacts, identify their sources, and present approaches both to avoid incorrect registrations and to confirm correct final registrations.
Moreover, we find that assuming simple translational offsets is an excellent first order approximation which requires little sacrifice in the final quality of the reconstructed images.
Using the acquisition and registration technique described here, we demonstrate cryo-STEM imaging with 0.72 \AA~information transfer, and clearly distinct atomic columns of disparate $Z$ values at $< 2$~\AA~spacing.



\section{Theory}
\label{S:theory}

\subsection{Referenceless correlation}

\begin{figure}[!htbp]
  \includegraphics[width=\linewidth]{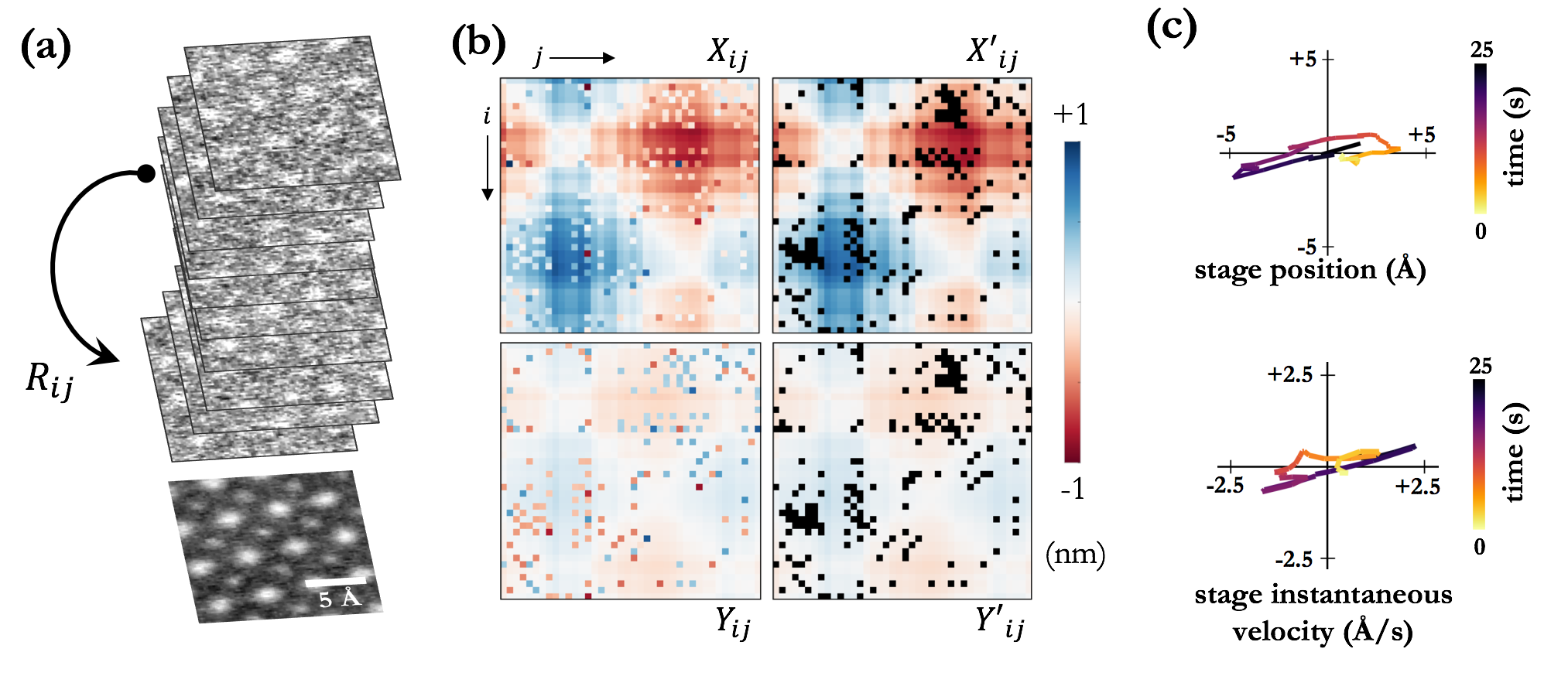}
    \caption{\textbf{Referenceless cryo-STEM image correlation.}
    (\textbf{a}) All possible image pairs in a stack of fast acquisition cryo-STEM images are cross correlated to determine their relative shifts.
    (\textbf{b}) The shift matrices $\mathbf{R}_{ij} = X_{ij}\mathbf{\hat{x}} + Y_{ij}\mathbf{\hat{y}}$ visualize the calculated shifts between all image pairs $(i,j)$ (\textit{left}), from which the optimum global image shifts may be calculated.
    The smoothly varying background encodes the stage movement during acquisition, while the abberant pixels are incorrect correlations, resulting from the low-SNR of cryo-STEM imaging.
    False correlations can then be identified (\textit{right}) and corrected.
    (\textbf{c}) Characterization of stage drift during acquisition can be extracted directly from the shift matrices, including both the stage position (\textit{top}) and instantaneous velocity (\textit{bottom}) as a function of time.
    In this case the stage drifted preferentially along the $x$-direction, changed direction multiple times, and had a maximum velocity of $\sim$2~\AA/s during the 25.2 s stack acquisition.}
  \label{F:registration}
\end{figure}

The cross correlation of real valued functions $f$ and $g$
\begin{equation}\label{E:cc_def}
    (f\star g)(x) \equiv \int^{\infty}_{-\infty} f(y)g(x+y)dy
\end{equation}
is interpretable as the overlap of $f$ and $g$ given some relative shift $x$.
For a pair of identical, translationally offset images, the correct shift for an optimum registration is therefore given by the value of the argument $x$ which maximizes the cross correlation -- see, e.g., \cite{modersitzki2004}.
Typically, all images in a series are registered to a single image.
Iterative schemes may then re-register using the output averaged image as a reference one or more times \cite{Berkels2014,Frank1981}.
In low-SNR data, a single incorrect cross correlation can spoil an entire reconstruction.
An ad hoc approach may be employed, whereby incorrectly registered images are discarded, or various images are tested as the reference.
However, this approach may discard useful data, and requires significant and subjective user input.
Moreover, incorrect correlations can introduce subtle artifacts which can be difficult to detect, but result in spurious analysis - see Fig.~\ref{F:data} and the associated text in the Results section.

The approach here is to correlate all pairs of images.
This has numerous advantages.
First, it is possible to calculate the optimum image shifts based on more complete information of all relative image shifts.
Second, construction of the matrix $\mathbf{R}_{ij}$ of shifts between all image pairs $(i,j)$ allows straightforward determination of incorrect correlations, which may then be corrected.
Additionally, characterization of stage stability and drift is then readily retreivable with little additional effort.

For a stack of $N$ image frames, we calculate the relative shifts $\mathbf{R}_{ij} = X_{ij}\mathbf{\hat{x}} + Y_{ij}\mathbf{\hat{y}}$ between all image pairs $(i,j)$ from their cross-correlations, shown in Fig.~\ref{F:registration}a.
Fig.~\ref{F:registration}b shows $X_{ij}$ and $Y_{ij}$ (\textit{left}) for an atomic resolution, experimental cryo-STEM dataset with $N=40$, described in more detail in Section~\ref{S:data}.
Here, the $i$'th row describes all measured shifts relative to the $i$'th image.
Assuming all images are related by a simple translational offset determined by the stage positions $(\mathbf{r}_i, \mathbf{r}_j)$ during the two acquisitions, the ideal shift matrix is $\mathbf{R}_{ij} = \mathbf{r}_j - \mathbf{r}_i$, which is manifestly skew-symmetric ($\mathbf{R}_{ji} = -\mathbf{R}_{ij}$).
Notably, noise sources with characteristic frequencies greater than the frame time cannot be directly corrected under such a simplifying assumption;
however, a range a high frequency noise sources, such as scan noise, can be averaged out, and we find that the combination of fast acquisitions ($\lesssim 1$ s) and sufficiently many ($\gtrsim 25$) frames to effectively average over higher frequency noise sources results in excellent image reconstructions.
Other high frequency noise sources include nitrogen bubbling, which often results in significant image distortions during a small number of frames.
This generally results in many outliers in a single row of the shift matrix, therefore the corresponding image frames can be subsequently excluded.

By eye, it is apparent that the shift matrices (e.g. Fig.~\ref{F:registration}b) contain smoothly varying backgrounds, and a number of aberrant pixels.
$X'_{ij}$ and $Y'_{ij}$ show the shift matrices with a mask indicating identified outliers - further discussion of outliers is in Section~\ref{S:outliers}. 
The smooth backgrounds directly encode the stage movement during image acquisition.
For example, in Fig.~\ref{F:registration}b, examining the trend in the smooth background of $X_{ij}$ from left to right, we see that the stage began by drifting in the positive $x$-direction over the first $\sim$10 frames, drifted back in the negative $x$-direction for another $\sim$18 frames, then drifted in the positive direction again for the final $\sim$12 frames.
This trend is apparent in any given row of $X_{ij}$; the values in two rows should in principle (absent noise) describe the same stage movements, with the origin shifted to a different reference image.

We determine the optimum shifts by calculating the most likely stage position during each of the $N$ acquisitions, given the set of all measured relative image shifts.
The problem is analytically tractable, and the optimum shift $\mathbf{r}_i$ for image $i$ is
\begin{equation}\label{E:shifts}
    \mathbf{r}_i = \frac{1}{N} \sum_j \mathbf{R}_{ij}
\end{equation}
The result is intuitively satisfactory: the optimum shift for image $i$ is simply the mean of the $i$'th row of $\mathbf{R}_{ij}$.
Derivation of Eq.~\ref{E:shifts} is found in \ref{A:shifts}.

Because the optimal shifts correspond to stage positions, combining this information with the known frame acquisition time leads to a description of stage drift during image acquisition.
Plotting the stage positions in time (Fig.\ref{F:registration}c, \textit{top}) shows the stage drift changed directions twice, drifted preferentially in the $x$-direction, and the stage position spanned a distance of $\sim$1 nm over the $\sim$20 s of imaging, consistent with the direct observations of $X_{ij}$ and $Y_{ij}$.
The instantaneous velocity of the stage was calculated using a St\"{o}rmer-Verlet algorithm after applying a small smoothing filter to the stage positions, and is shown in Fig.~\ref{F:registration}c, (\textit{bottom}) \cite{allen1989}.
In this dataset, the maximum instantaneous drift velocity magnitude was $\sim$2 \AA/s. 
In addition to the two most obvious direction changes observable from the stage positions, we find several smaller kinks in the stage's velocity vector.
While here the stage movement qualitatively appears well described by a (possibly biased) random walk, stage drift varies both quantitatively and qualitatively as a function of holder, the presense of liquid nitrogen for cooling, and room environment.

\subsection{Unit cell jumps and sampling error}

In atomic resolution data, incorrect correlations are frequently the result of unit cell jumps, in which the two images are shifted by some linear combination of the crystal lattice vectors.
For a truly perfect crystal, unit cell hops are meaningless.
In real data, we are frequently interested in small perturbations from ideal structures, which will be smoothed out by unit cell hops -- see Fig.~\ref{F:data} in the Results section.

\begin{figure}[!htbp]
  \includegraphics[width=0.6\linewidth]{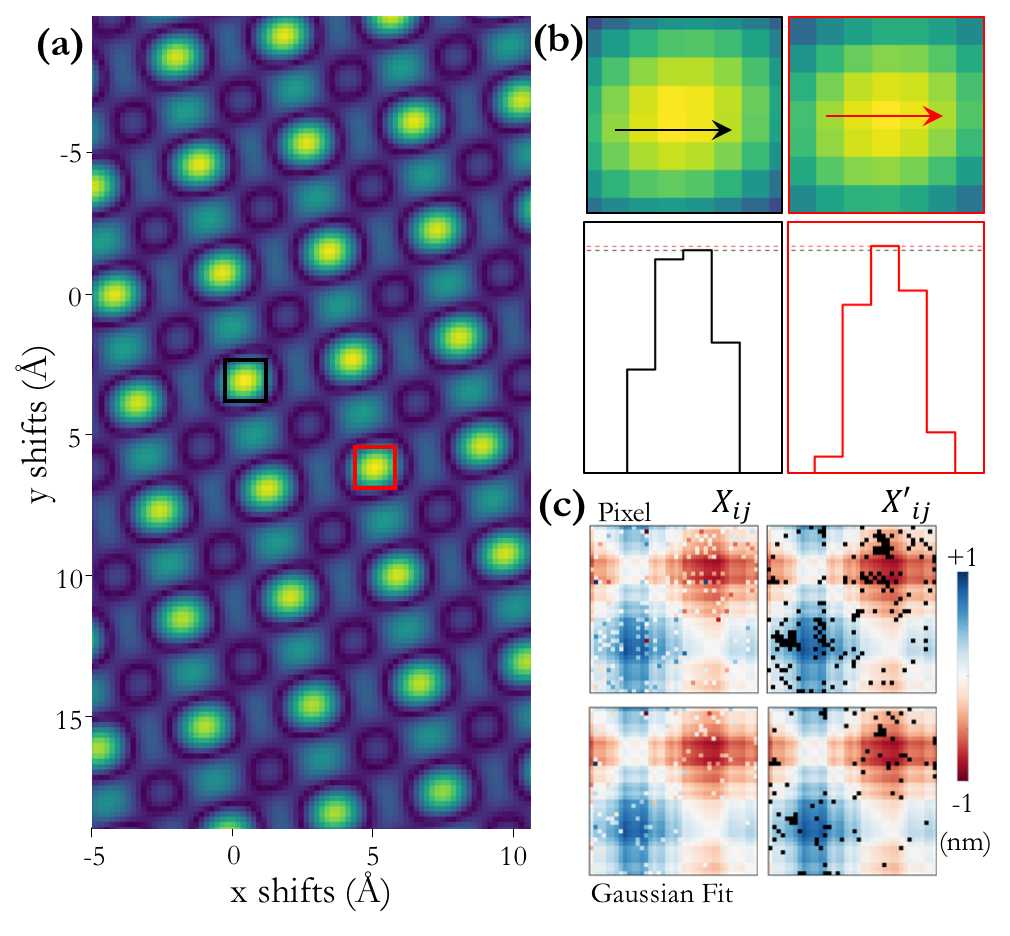}
  \centering
    \caption{\textbf{Unit cell jumps and sampling error.}
    (\textbf{a}) The cross correlation between a pair of atomic resolution cryo-STEM images contains many local maxima, corresponding to crystal lattice vector offsets between the images.
    (\textbf{b}) In this case, the intensity of the correct maximum (\textit{left, black box}) has been distributed among four adjacent pixels, while the intensity of the incorrect maximum (\textit{right, red box}) falls primarily in a single, central pixel.
    As a result, the brightest pixel in the cross correlation is found in the red region, shown in the line profiles (\textit{bottom}).
    (\textbf{c}) The $x$-shift matrix obtained by identifying the maximum pixel in each cross correlation contains incorrectly identified shifts (\textit{top}), many of which result from the sampling-induced unit cell hops seen here.
    Calculating the same matrix by identifying the cross correlation maximum with gaussian fits to the brightest several local maxima removes 50\% of the erroneous matrix elements (\textit{bottom}).
    This additionally yields the relative image shifts with subpixel resolution.}
  \label{F:sampling}
\end{figure}

Fig.~\ref{F:sampling} shows how correlations often fail, illustrated using a particular failed correlation in the dataset discussed in Fig.~\ref{F:registration}.
The cross correlation (Fig.~\ref{F:sampling}a) contains many local maxima, corresponding to unit cell shifts between the two images.
The correct maximum, corresponding to the true relative stage shifts, is indicated in black (Fig.~\ref{F:sampling}b, \textit{left}).
However, the maximum pixel in the correlation is instead in the region indicated in red (Fig.~\ref{F:sampling}b, \textit{right}).
The source of the error here is sampling.
The center of the peak in the black region is located between the central four pixels, while the maximum in the red region is located almost exactly at the center of a single pixel.
The result is that although the black region contains the correct shift, the brightest pixel in the image is in the red region, evident in line cuts shown.

Identifying the relative shift between these two images using the maximum pixel in the cross correlation will therefore result in a unit cell hop.
One solution is to increase the pixel density per atom during data acquisition.
However, this is not always possible, for example when fast acquisition times and/or large fields of view are required.
Here, we determine the correct image shifts by first identifying the 3--5 local maximima containing the brightest few pixels in the cross correlation.
We then fit two dimensional Gaussians to each of these local maxima, and identify the correct shift as the global maximum among the resulting fit curves.
Fig.~\ref{F:sampling}c shows a 50\% reduction in incorrect correlations using this approach.
Fitting additionally allows identification of the relative image shifts with subpixel resolution.
Note, however, that because the global image shifts are determined from all relative shifts via Eq.~\ref{E:shifts}, calculating cross correlation maxima with pixel resolution still yields subpixel global shifts, with pixelation contributing error $\approxpm 0.5/\sqrt{N}$.

Unit-cell jumps may result from additional sources, discussed further in the Results section, therefore some outliers may remain after accounting for sampling problems.
However, due to the information surplus, image reconstruction remains achievable by identifying incorrect correlations, then calculating the ideal $\mathbf{R}_{ij}$ matrix in best agreement with the physically consistent shift measurements, discussed next.

\subsection{Incorrect correlation handling}
\label{S:outliers}

$X_{ij}$ and $Y_{ij}$ are $N\times N$ matrices, comprised of $N(N-1)/2$ measurements of relative image shifts, with the remaining $(N+1)N/2$ elements determined by skew-symmetry.
However, to first order, these measurements correspond to only $N$ distinct physical quantities - i.e. stage positions.
The redundancy corresponds to the physical requirement that the shifts preserve additive transitivity ($\mathbf{R}_{ij} + \mathbf{R}_{jk} = \mathbf{R}_{ik}$).
In practice, this condition rarely holds for imperfect, experimental data.
In such cases, the information surplus may be leveraged both to identify and remove incorrect correlations, and to determine the optimal shifts in spite of missing shift matrix elements.

Outliers are determined by identifying elements of $\mathbf{R}_{ij}$ which break additive transitivity.
For each element $(i,k)$, there are $2^{N-2}-1$ equations of the form
\begin{equation}\label{E:paths}
    \mathbf{R}_{ik} = \mathbf{R}_{ij_1} + \mathbf{R}_{j_1j_2} + \ldots + \mathbf{R}_{j_nk}
\end{equation}
which must hold for an ideal shift matrix, for some integer $n \in [1,N-2]$.
Physically, each corresponds to a stage trajectory involving a subset of the imaging positions.
As a practical matter, it is unneccesary and computationally prohibitive to use all such paths to identify and account for outliers, therefore we make use of a small subset for each matrix element.
Typically, $\sim$5 relationships are sufficient to ensure consistency.
The equations selected to evaluate the physical consistency of each shift matrix element are chosen to preference more trustworthy measurements.
In particular, our implementation prioritizes:
\begin{enumerate}
\item registrations $\mathbf{R}_{j_1j_2}$ minimizing $\lvert j_2 - j_1\rvert$, corresponding to shorter times between image acquisitions, and
\item paths consisting of stage positions $j_0$ for which $i<j_0<k$, corresponding to evaluating the fidelity of $\mathbf{R}_{ij}$ using events that occured in the time between acquisitions $i$ and $j$.
\end{enumerate}

Outliers are identified by calculating the mean absolute difference between the left and right sides of Eq.~(\ref{E:paths}) over all selected paths for each matrix element, then performing a simple threshhold.
Our code includes several additional outlier detection approaches, including comparison to nearest neighbor elements, as well as deviation from a background fitting function.
Incorrect $\mathbf{R}_{ij}$ elements are then replaced with values that best enforce transitivity, using Eqs.~(\ref{E:paths}) over paths containing only correct correlations -- see Fig.~\ref{F:informationredundancy}.
Finally, optimal shifts are determined from Eq.~\ref{E:shifts}.


\section{Results}
\label{S:results}

Cryo-STEM experiments were performed on an aberration corrected FEI Titan Themis 300 operating at 300 kV, in conjunction with a side entry, double tilt liquid nitrogen holder (Gatan 636).
To maintain good insultation between the cryogen and the room environment we baked the dewar vacuum at 100 C for 12 hours prior to imaging, and to minumize sample drift we let the holder settle after cooling in the microscope for at least 2 hours.
Samples were prepared by focused ion beam lift-out, imaged with 30 mrad and 21.4 mrad convergence semiangles for Bi$_{1-x}$Sr$_{x-y}$Ca$_{y}$MnO$_3$ and Nb$_3$C$_8$ crystals, respectively, and in both cases with HAADF detector inner and outer angles of 68 and 340 mrad, respectively.

\subsection{Atomic resolution cryo-STEM image registration}
\label{S:data}

\begin{figure}[!htbp]
  \includegraphics[width=\linewidth]{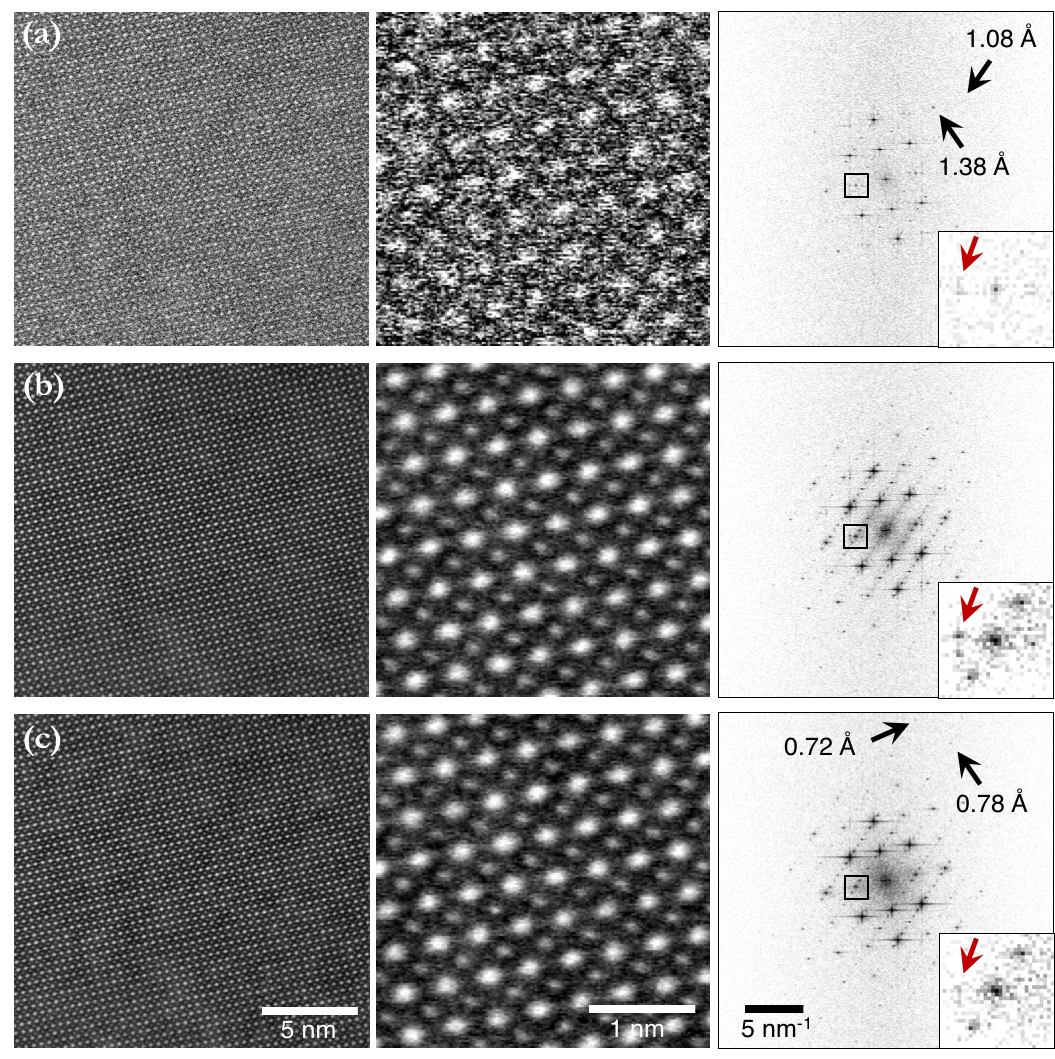}
    \caption{\textbf{Avoiding subtle artifacts in atomic resolution cryo-STEM image registration.}
    (\textbf{a}) A single fast acquisition (0.63 s) frame from a 40 image series of Bi$_{1-x}$Sr$_{x-y}$Ca$_{y}$MnO$_3$ (BSCMO) imaged under liquid nitrogen cooling.
    In the full field of view (\textit{left}) and zoom-in (\textit{middle}) the brighter A-sites of the perovskite lattice are identifiable but noisy, while the dimmer B-sites cannot be clearly distinguished.
    The FFT (\textit{right}) shows clear information transfer to 1.38~\AA, and very weak reflections at 1.08~\AA.
    (\textbf{b}) Registering all frames to a single reference frame and averaging significantly enhances the SNR, however, this registration is subtly flawed due to unit cell jumps, and any subsequent analysis would be untrustworthy. 
    (\textbf{c}) The optimum registration, obtained by determining the optimal shifts from all image pairs, shows information transfer to 0.72~\AA.
    The incorrect registration differs in several ways from the optimum registration.
    First, variation in the A-site intensity is apparent in the correctly registered full field of view image, and corresponds to cation disorder in the sample, but has been averaged out in the incorrect registration.
    Second, artificial peaks in the incorrectly registered FFT (\textit{right inset, red arrow}) are suppressed in the optimally registered data.
    Noise of known origin (overlapped cross-hatch streaking from two Bragg peaks) at this frequency may have contributed to the failed registrations.
    Finally, BSCMO supports a charge density wave state which results in periodic, picometer scale shifts of the atomic columns, with a wavelength of $\sim$3 unit cells.
    Unit cell hops have therefore introduced diagonal streaks in the diffuse background of the incorrect registration's FFT, while the diffuse background of the optimally registered FFT decays monotonically.}
  \label{F:data}
\end{figure}

Figure~\ref{F:data} shows STEM data of the manganite Bi$_{1-x}$Sr$_{x-y}$Ca$_{y}$MnO$_3$ (BSCMO) where $x=0.65$ and $y=0.47$.
The brighter atomic columns are the Bi/Sr/Ca atoms occuping the A-sites of a perovskite lattice,  while the dimmer atomic columns are the manganese B-sites.
The data was acquired under liquid nitrogen cooling, with a 0.5 $\mu$s pixel dwell time and 1024x1024 pixel frames. 
A series of 40 frames was acquired over 25.2 s.
Figure~\ref{F:data}a shows the full field of view (\textit{left}), a zoomed in region (\textit{middle}), and the FFT (\textit{right}) from a single frame.
While the frame contains clear periodicity, the SNR is poor.
The brighter A-sites are apparent but noisy, while the dimmer B-sites are not clearly distinguishable from background noise.

Figures~\ref{F:data}b,c both show the image stack after registration and averaging.
Figure~\ref{F:data}b was registered to a single reference image, and several frames were incorrectly registered with discrete unit-cell jumps, while Fig.~\ref{F:data} was registered using all image correlation pairs to determine the optimum shifts and exclude any unit-cell jumps.
As a result, the full field of view in Fig.~\ref{F:data}b shows a nearly perfectly smooth lattice, in contrast to the dappled contrast apparent in Fig.~\ref{F:data}c.
This local contrast variation corresponds to cation disorder of the Bi/Sr/Ca atoms (atomic numbers 83/38/20), which form a solid solution in the A-sublattice.
The more perfect appearance of the incorrect registration upon visual inspection thus reflects averaging out of real sample features.

The incorrect registration has introduced an additional, more subtle artifact into the averaged image.
BSCMO supports a charge density wave state, and in this data the atomic lattice sites are each displaced from their ideal positions by $\sim$7-11 pm in a periodic pattern with a wavelength of $~3$ unit cells \cite{savitzky2017,elbaggari2017}.
These periodic lattice displacements (PLDs) are apparent from the two satellite peaks adorning each Bragg peak in the FFTs \cite{Hovden2016}.
Averaging image frames with an incorrect unit cell jump results in averaging the PLD with itself plus a phase offset.
Tracking the atomic displacements in this image to observe the local behavior of the PLD may therefore give incorrect, misleading results.
In the FFT (Fig.~\ref{F:data}b, \textit{right}), this manifests as subtle stripes in the diffuse background along the direction of the satellite peaks.
In the optimally registered FFT (Fig.~\ref{F:data}), the diffuse background decreases monotonically with $\lvert\mathbf{k}\rvert$, as expected for this sample.
Artifacts in the FFT diffuse background are illustrated more prominently in Supplementary Figures S1-2.

In this particular dataset, it is possible to identify a possible source of the failed correlations.
The highlighted Bragg peak (Fig.~\ref{F:data}, \textit{right insets}) appears slightly different in each of the three FFTs shown.
In the optimally registered FFT (Fig.~\ref{F:data}c), the Bragg peak and two satellite peaks are visible, with very little intensity present at the indicated point (\textit{red arrow}).
In the incorrectly registered FFT (Fig.~\ref{F:data}b), there are additional peaks present at the indicated point, which do not correspond to any physical periodicity in BSCMO, and are absent both in other similar STEM datasets as well as in diffraction.
In the single frame FFT (Fig.~\ref{F:data}a), there is a small amount of noise present at this location, which happens to fall at the intersection of the cross-hatch streaks from the two nearest Bragg peaks.
We surmise that the confluence of this frequency space noise pattern with a unit cell jump may have led to the incorrect registrations in this dataset.
The failed registration shows enhanced noise at this point, creating the artificial spots in the FFT of Fig.~\ref{F:data}b, while the optimally registered image suppresses this noise.
For simple Poisson noise, the expected SNR improvement scales with $\sqrt{N}$, and here we find that for the optimally average data in Fig.~\ref{F:data}, $\text{SNR}_{ave}/\text{SNR}_{frame}=0.79\sqrt{N}$ with $N=40$.\footnote{SNR estimates are obtained by approximating the noise as $I(r_{ij}) - I(r_{ij})*\mathcal{N}_{\sigma=2}$, where $I(r_{ij})$ is the image intensity at pixel $r_{ij}$, $*$ is a convolution, and $\mathcal{N}_{\sigma=2}$ is a gaussian kernel with a 2 pixel standard deviation.}
However, the functional form of the SNR improvement with $N$ does not obey a simple power law, suggesting more complex noise sources (see Supplementary Figure S3).
The FFT of the optimally registered and averaged image shows information transfer to $0.72 \AA$ (\textit{black arrows}).

\subsection{Fourier weighting to minimize registration errors}

As a practical matter, cross correlations are typically performed in reciprocal space due to the computational efficiency of the fast Fourier transform.
It is therefore convenient to write Eq.~(\ref{E:cc_def}) in the form suggested by the cross correlation theorem:
\begin{equation}\label{E:cc}
    (f\star g)(x) = \mathcal{F}^{-1}\big( (\mathcal{F}f)^{\dagger}(\mathcal{F}g) \big)
\end{equation}
where $\mathcal{F}$ and $\mathcal{F}^{-1}$ are the Fourier transform and its inverse, respectively, and $^\dagger$ indicates the complex conjugate.

In low-SNR data, noise can dominate calculation of Eq.~\ref{E:cc}.
Obtaining a correct correlation thus necessitates consideration of where the most important information resides in frequency space.
We therefore apply a weighting function $w(\mathbf{k})$ in Fourier space before performing the inverse transform:
\begin{equation}\label{E:mask}
    (f\star g)(x) = \mathcal{F}^{-1}\big( w(\mathbf{k}) (\mathcal{F}f)^{\dagger}(\mathcal{F}g) \big)
\end{equation}
For higher SNR data, it is often sufficient to choose a low pass or bandpass filter for $w(\mathbf{k})$, with a high frequency cutoff at the information limit of the data to exclude high frequency noise, or to exclude Fourier weighting altogether.
For low-SNR data, and highly translationally symmetric lattice images in particular, determining a weighting function which will best extract the true image shifts requires more careful inquest.

\begin{figure}[!htbp]
  \includegraphics[width=0.75\linewidth]{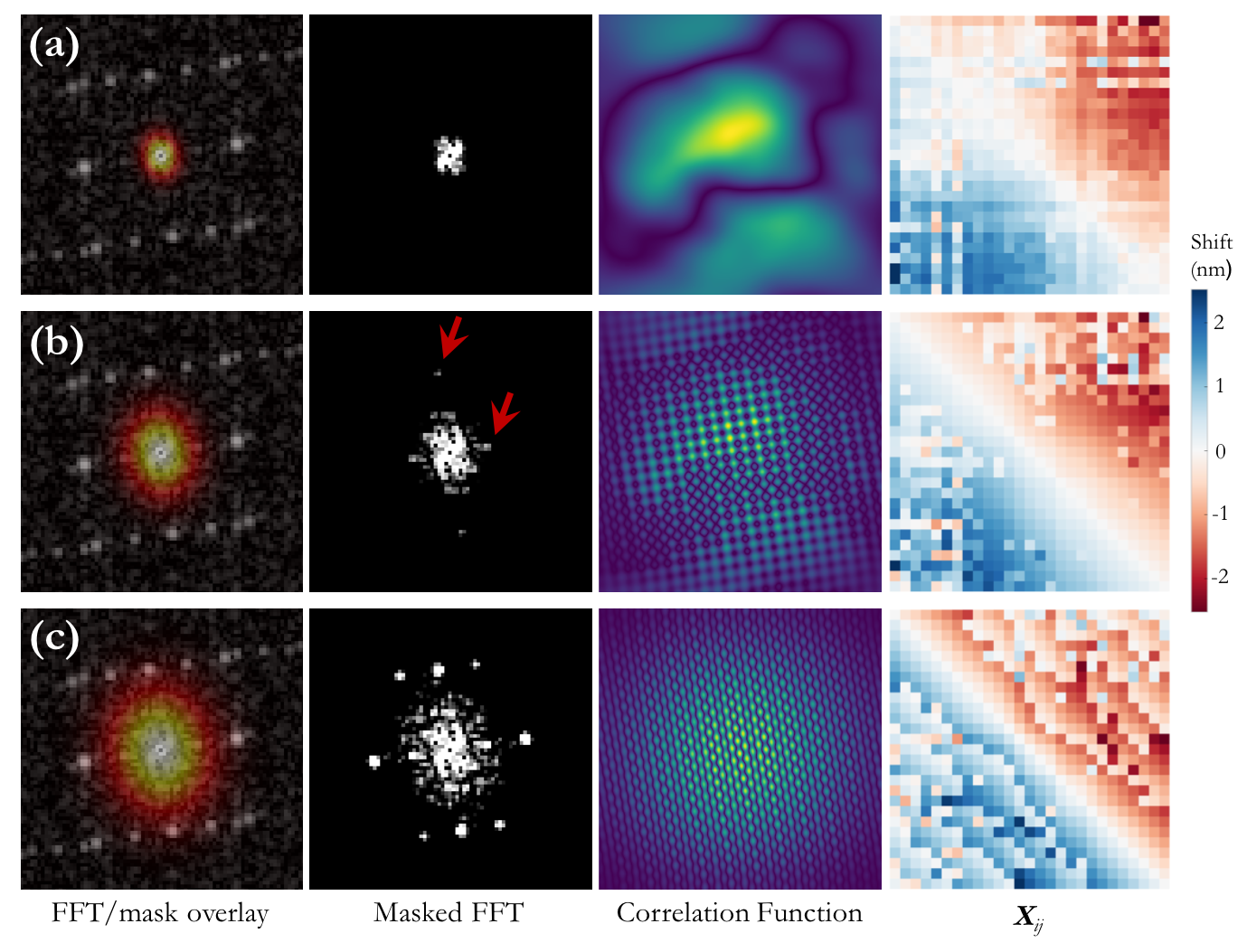}
    \centering
    \caption{\textbf{Weighting information in Fourier space.}
    (\textbf{a}) Applying a Fourier mask with a high frequency cutoff below the primitive reciprocal lattice vectors (\textit{left, center left}) removes all lattice information from the cross correlation (\textit{center right}).
    The cross correlation is therefore smooth, and correctly identifies the general region of the global maximum, however lacks any precision in locating the maximum within that region, resulting in a coarse $\mathbf{R}_{ij}$ structure (\textit{right}, $X_{ij}$ shown).
    (\textbf{b}) A Fourier mask with a high frequency cutoff just above the primitive reciprocal lattice vectors (\textit{left, center left}) contains the many local maxima corresponding to the atomic lattice, while simultaneously identifying the region of the global maximum by significantly weighting low frequencies (\textit{center right}).
    The resulting image shifts are far more precise than those in (a), manifesting as smoother $\mathbf{R}_{ij}$ matrices in areas without unit cell jumps (\textit{right}).
    (\textbf{c}) A Fourier mask with a high frequency cutoff well above the primitive reciprocal lattice vectors (\textit{left, center left}) results in a cross correlation which is dominated by lattice information (\textit{center right}), with the region of the global maximum entirely obscured due to the combination of low-SNR and high translational symmetry.
    The resulting $\mathbf{R}_{ij}$ matrices contain many unit cell hops, identifying the local maximum which minimizes the relative shift between the image pair.}
  \label{F:maskeffects}
\end{figure}

\begin{figure}[!htbp]
  \includegraphics[width=\linewidth]{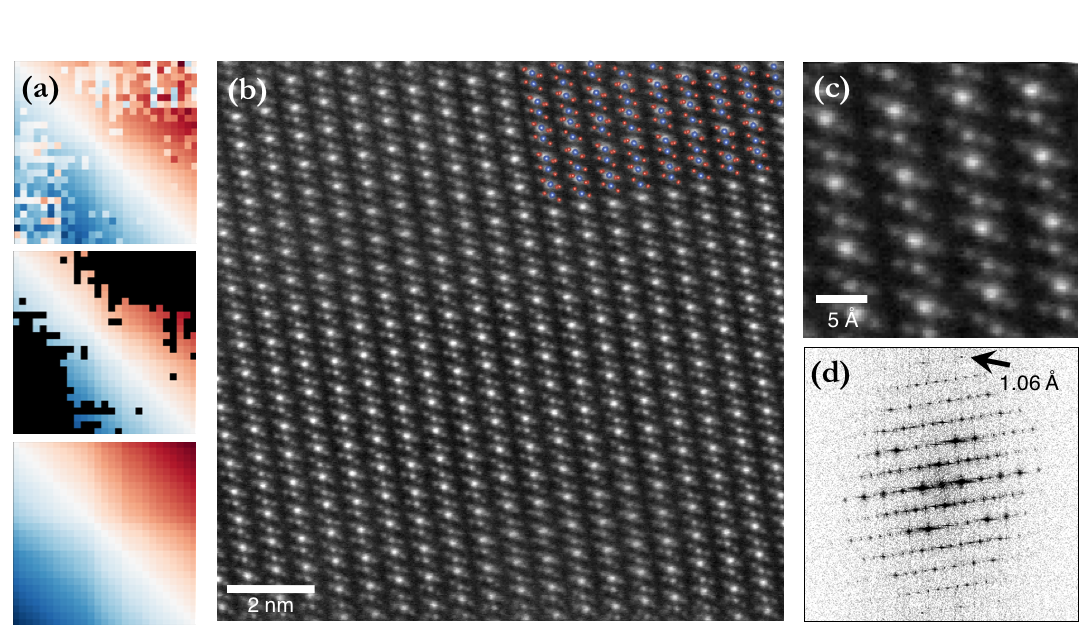}
    \caption{\textbf{Information redundancy and outlier handling in cryo-STEM image registration.}
    (\textbf{a}) The $\mathbf{R}_{ij}$ matrices often contain incorrect correlations, particularly in image pairs acquired a long duration of time apart, i.e., the upper right and lower left corners of $\mathbf{R}_{ij}$ (\textit{top}).
    Calculating the deviation of each matrix element from perfect transitivity with Eqs.~\ref{E:paths} and thresholding identifies incorrect correlations (\textit{middle}).
    Physically consistent values for missing elements can then be determined using transitivity, and the optimal shifts calculated with Eq.~\ref{E:shifts} (\textit{bottom}).
    (\textbf{b}) The resulting registered and averaged 27 image stack of 512x512 pixel images and 0.58 s frame times of the layered material Nb$_3$Cl$_8$, imaged in cross section.
    (\textbf{c}) A zoomed-in region of (b) illustrates the benefit of achieving high-SNR STEM images -- every Nb and Cl column is clearly distinguishable, in spite of their disparate atomic numbers ($Z_\text{Nb}=41$, $Z_\text{Cl}=17$) and projected interatomic spacings of 1.66~\AA~to 2.07~\AA.
    (\textbf{d}) The FFT shows information transfer to 1.06~\AA.}
  \label{F:informationredundancy}
\end{figure}

Figs.~\ref{F:maskeffects} and \ref{F:informationredundancy} examine cryo-STEM data of the layered material Nb$_3$Cl$_8$ \cite{Sheckelton2017}.
A series of 27 frames was acquired over 15.5 s, with a 2 $\mu$s dwell time and 512x512 pixel frames, under liquid nitrogen cooling.
Fig.~\ref{F:maskeffects} shows the effect of varying the Fourier mask $w(\mathbf{k})$.
In each case, $w(\mathbf{k})$ is an anisotropic gaussian with principle axes oriented along the reciprocal lattice basis vectors $\mathbf{b}_i$.
In Fig.~\ref{F:maskeffects}a, $w(\lvert\mathbf{k}\rvert > k_\text{max})\ll1$ for some $k_\text{max} < \lvert\mathbf{b}_i \rvert$.
Thus, the lattice has been entirely discarded, and only low frequency information has been retained, apparent from the FFT/$w(\mathbf{k})$ overlay (\textit{left}) as well as the masked FFT (\textit{center left}) from a representative frame, which no longer displays any Bragg peaks.
The resulting correlation function (\textit{center right}) contains a single global maximum, and lacks the many local maxima seen in Fig.~\ref{F:sampling}.
By registering signal components of size scales larger than the unit cell, this mask avoids unit cell hops and finds the approximate region of the correct shift.
The result is far fewer large, discontinuous jumps in the shift matrix $X_{ij}$ compared to the other two Fourier masks shown (\textit{right}).
However, the precise location of the correct shift within this approximate region cannot be determined without lattice information, resulting in the coarse structure of this $X_{ij}$.

Fig.~\ref{F:maskeffects}c shows a larger $w(\mathbf{k})$, with $k_\text{max} > \lvert\mathbf{b}_i\rvert$, incorporating several Bragg peaks into the registration (\textit{left, center left}).
The resulting cross correlation contains many local maxima (\textit{center right}).
However, the SNR in this data is low enough that the low frequency information, which identified the approximate region of the globally correct shift in Fig.~\ref{F:maskeffects}a, is largely dominated by the lattice in this case.
The result is that unit cell jumps are far more likely with this mask.
Examining $X_{ij}$ (\textit{right}) confirms that this is the case.
Nearby shift matrix elements are generally smooth and continuous, such as those near the central diagonal, indicating that registrations between images taken a short time apart tend to be correct here.
However, several discontinuous steps occuring with increasing distance from the central diagonal suggest that as the time between image acquisitions grows, unit cell jumps become increasingly likely.

Fig.~\ref{F:maskeffects}b shows a $w(\mathbf{k})$ with $k_\text{max}\approx\lvert\mathbf{b}_i\rvert$ (\textit{left}).
This mask heavily weights low frequencies, but still includes some atomic lattice information (\textit{center left, red arrows}).
The slow, smooth background comprising the cross correlation in Fig.~\ref{F:maskeffects}a is clearly present here (\textit{center right}).
At the same time, the many local maxima corresponding to the atomic lattice are preserved.
There is thus sufficient low frequency information here to identify the approximate region of the global cross correlation maximum, and sufficient lattice information to lock the maximum correctly to the atomic lattice, without inducing unit cell hops to an incorrect local maximum.
The resulting shift matrix (\textit{right}) suffers neither from the many unit cell hops of the largest Fourier mask, nor from the coarse, imprecise structure of the smaller mask.
Alternatively, local minima may be avoided at the cost of additional computation time with heirarchical coarse-graining schemes \cite{Berkels2014,vural2013,alvarez1999scale}.

Our implementation includes various tunable Fourier mask options, and straighforward support for custom masks.
We find that results vary relatively little with the functional form of the apodization mask (Hann, Hamming, Blackman, Gaussian), but are far more sensitive to cutoff frequencies.
We additionally note that while circular masks ($w(\mathbf{k}) = w(\lvert\mathbf{k}\rvert)$) are generally sufficient for samples with rotational symmetries higher than C$_2$ in the projection direction (exactly or approximately), for anisotropic projected lattices non-circular mask shapes are recommended to avoid overweighting a particular lattice direction.

The optimized $X_{ij}$ shift matrix in Fig.~\ref{F:maskeffects}b, and the corresponding $Y_{ij}$, still contain many outliers, particularly in the upper right and lower left corners, representing registrations between image pairs separated by longer spans of time.
However, by enforcing information consistency as described above, there is ample information here to determine the optimal shifts.
The outlier handling procedure described in Section~\ref{S:outliers} is visualized for this dataset in Fig.~\ref{F:informationredundancy}a.
From the initial $\mathbf{R}_{ij}$ matrices (\textit{top}, $X_{ij}$ shown), the deviation of each registration from perfect additive transitivity is calculated following Eqs.~\ref{E:paths}.
Threshholding (\textit{middle}) effectively identifies incorrect correlations.
Calculating best estimates for these missing matrix elements by transitivity yields smooth shift matrices which correspond to an approximately constant stage drift during this image series acquisition.

The resulting registered and averaged image is shown in Fig.~\ref{F:informationredundancy}b.
Here, we observe Nb$_3$Cl$_8$ near liquid nitrogen temperature in cross section.
The alternating light/dark pattern at the niobium sites (Fig.~\ref{F:informationredundancy}b, \textit{blue overlay}) occurs because the quasi-two dimentional layers are oriented such that alternating columns project through either one or two niobium atoms per unit cell of a kagome lattice \cite{Sheckelton2017}.
In this projection, the spacing between chlorine atoms and the most proximate niobium column ranges between 1.66~\AA~and 2.07~\AA.
Here, every individual Cl and Nb column (atomic numbers $Z_\text{Cl}=17$ and $Z_\text{Nb}=41$) are clearly identifiable in both the full field of view (Fig.~\ref{F:informationredundancy}b) and the zoom in (Fig.~\ref{F:informationredundancy}c). 
The FFT (Fig.~\ref{F:informationredundancy}d) shows information transfer to 1.06 \AA.



\section{Discussion}
\label{S:discussion}

Our approach differs from many other STEM image registration methods in that we assume all images in a series are related by simple translational offsets.
Strictly speaking, this is incorrect; real data is additionally plagued by nonlinear drift, scan noise, and imperfect scan coils.
An appropriately chosen continuous coordinate transformation between images can in principle account for image distortions which our approach will fail to correct.
Rotating scans have the additional benefit that the superior information transfer along the fast scan direction contributes more isotropically to the final reconstructed image \cite{Sang2014,Ophus2016}.

For low-SNR data, however, our rigid registration approach affords several advantages.
Most importantly, it allows registration of challenging datasets that otherwise might not be recoverable at all, and has proven vital for inherently difficult cryo-STEM data.
Moreover, simultaneously considering the correlations between all image pairs enables confirmation that the extracted shifts are physically consistent, thereby avoiding artifacts from unit cell jumps which can be difficult to detect, and lead to incorrect analysis.
Unlike rotated scan methods, the approach described here requires no specialized image acquisition code or procedures aside from fast acquisitions.
Unlike more computationally intensive optimization methods, our approach is fast - on a laptop computer we register 40 image stacks of 1024x1024 pixels in $\sim$1 minute, or of 512x512 pixels in a matter of seconds.
This is highly beneficial for low-SNR data, where it may be valuable to attempt registration with various parameter choices (e.g. Fourier masks).

Although a simple translation is far from a complete physical model of the differences between STEM image pairs, we find that for sufficiently fast scans (in which stage drift velocity can be considered approximately constant over a single frame) this first-order approximation is more than adequate.
We have demonstrated 0.72~\AA~information transfer and sufficient SNR to easily distinguish Cl and Nb columns at $<$2~\AA~spacing at cryogenic temperatures.
Impressive SNR improvements have been demonstrated previously with more sophisticated registration approaches, and are highly valuable for measurements requiring sub-picometer precision.
However, in low-SNR or highly translationally symmetric data, these methods may fail or require further refinement.
Moreover, rigid registration has been demonstrated to provide sufficient precision to observe a variety of physically important atomic distortions \cite{kimoto2010,savitzky2017,mundy2016}. 
Regardless of application, while post processing is an important step in achieving excellent final image quality, more crucial still is optimized instrumentation and data acquisition.
In cases where optimized data acquisition and noisy data are two sides of the same coin, such as in current cryo-STEM, our approach avoids pernicious sources of artifact using simple physical contraints.

\section{Conclusions}
\label{S:conclusions}

We have described an image registration approach which allows fast, accurate registration of difficult STEM datasets.
Combining registration of all possible image pairs with the simplifying assumption that the image pairs are related by translational offsets only yields a surfeit of information, allowing identification and correction of incorrect registrations which may plague low-SNR data.
Further, excess information allows direct confirmation that the final image shifts are physically consistent, avoiding the very real possibility of insidious artifacts due to unit cell jumps.
Unit cell jumps can be additionally minimized by accounting for sampling errors, and by judicious determination of the optimum weighting of frequency space information.
Despite the simplicity of our model, we find that the assumption of translational shifts only yield excellent results in terms of information transfer and SNR gains, which are more than adequate for many applications.
Developed for inherently challenging cryo-STEM data, our approach may also be useful for registering any low-SNR image stacks, particularly in cases of high translational symmetry.

\section*{Acknowledgements}\
\label{S:acknowledgements}

The authors thank Jim Sethna for useful conversations.
This work was supported by the National Sciences Foundation, through the PARADIM Materials Innovation Platform (DMR-1539918), the NSF GRFP (DGE-1144153, B.H.S.), the Center for Bright Beams graduate research fellowship (PHY-1549132, C.C.), and made use of the Cornell Center for Materials Research with funding from the NSF MRSEC program (DMR-1719875).
Additional support was provided by the Department of Defense Air Force Office of Scientific Research under award number FA 9550-16-1-0305.
The FEI Titan Themis 300 was acquired through NSF-MRI-1429155, with additional support from Cornell University, the Weill Institute and the Kavli Institute at Cornell.
The work at Rutgers was supported by the Gordon and Betty Moore Foundation's EPiQS Initiative through Grant GBMF4413 to the Rutgers Center for Emergent Materials.


\appendix

\section{Optimal Shifts}
\label{A:shifts}

The goal is to find a matrix of the form $\mathbf{R'}_{ij} = \mathbf{r}_j - \mathbf{r}_i$ which is closest to the measured relative image shifts $\mathbf{R}_{ij}$.
Thus, we are interested in finding
\begin{equation*}
    \text{min}_{\{\mathbf{r}_i\} } \lVert \mathbf{R}_{ij} + \mathbf{r}_i - \mathbf{r}_j \rVert ^{2}
\end{equation*}
Direct computation yields
\begin{align*}
    0 &= \frac{\partial}{\partial \mathbf{r}_k} \sum_{ij} \left (\mathbf{R}_{ij} + \mathbf{r}_i - \mathbf{r}_j \right )^2 \\
      &= \sum_{ij} \left ( \mathbf{R}_{ij}\delta_{ik} - \mathbf{R}_{ij}\delta_{jk} - \mathbf{r}_i\delta_{jk} - \mathbf{r}_j\delta_{ik} + \mathbf{r}_i\delta_{ik} + \mathbf{r}_j\delta_{jk} \right) \\
    \mathbf{r}_k &= -\frac{1}{N}\sum_j \mathbf{R}_{kj} + \frac{1}{N}\sum_j\mathbf{r}_j
\end{align*}
where $\delta_{ij}$ is the Kronecker delta.
Choosing $\frac{1}{N}\sum_j\mathbf{r}_j$ as the origin yields Eq.~\ref{E:shifts}.

\section*{References}
\label{S:references}

\bibliographystyle{elsarticle-num} 
\bibliography{references}

\begin{thebibliography}{10}
\expandafter\ifx\csname url\endcsname\relax
  \def\url#1{\texttt{#1}}\fi
\expandafter\ifx\csname urlprefix\endcsname\relax\def\urlprefix{URL }\fi
\expandafter\ifx\csname href\endcsname\relax
  \def\href#1#2{#2} \def\path#1{#1}\fi

\bibitem{haider2000upper}
M.~Haider, S.~Uhlemann, J.~Zach,
  \href{http://www.sciencedirect.com/science/article/pii/S0304399199001941}{Upper
  limits for the residual aberrations of a high-resolution aberration-corrected
  stem}, Ultramicroscopy 81~(3) (2000) 163 -- 175.
\newblock \href
  {http://dx.doi.org/https://doi.org/10.1016/S0304-3991(99)00194-1}
  {\path{doi:https://doi.org/10.1016/S0304-3991(99)00194-1}}.
\newline\urlprefix\url{http://www.sciencedirect.com/science/article/pii/S0304399199001941}

\bibitem{Batson2002}
P.~E. Batson, N.~Dellby, O.~L. Krivanek, {Sub-{\aa}ngstrom resolution using
  aberration corrected electron optics.}, Nature 418~(6898) (2002) 617--620.
\newblock \href {http://dx.doi.org/10.1038/nature01058}
  {\path{doi:10.1038/nature01058}}.

\bibitem{tanaka2008present}
N.~Tanaka, \href{http://stacks.iop.org/1468-6996/9/i=1/a=014111}{Present status
  and future prospects of spherical aberration corrected tem/stem for study of
  nanomaterials}, Science and Technology of Advanced Materials 9~(1) (2008)
  014111.
\newline\urlprefix\url{http://stacks.iop.org/1468-6996/9/i=1/a=014111}

\bibitem{Muller2006}
D.~A. Muller, E.~J. Kirkland, M.~G. Thomas, J.~L. Grazul, L.~Fitting,
  M.~Weyland, {Room design for high-performance electron microscopy},
  Ultramicroscopy 106~(11-12 SPEC. ISS.) (2006) 1033--1040.
\newblock \href {http://dx.doi.org/10.1016/j.ultramic.2006.04.017}
  {\path{doi:10.1016/j.ultramic.2006.04.017}}.

\bibitem{Sang2014}
X.~Sang, J.~M. Lebeau,
  \href{http://dx.doi.org/10.1016/j.ultramic.2013.12.004}{{Revolving scanning
  transmission electron microscopy : Correcting sample drift distortion without
  prior knowledge}}, Ultramicroscopy 138 (2014) 28--35.
\newblock \href {http://dx.doi.org/10.1016/j.ultramic.2013.12.004}
  {\path{doi:10.1016/j.ultramic.2013.12.004}}.
\newline\urlprefix\url{http://dx.doi.org/10.1016/j.ultramic.2013.12.004}

\bibitem{Berkels2014}
B.~Berkels, P.~Binev, D.~A. Blom, W.~Dahmen, R.~C. Sharpley, T.~Vogt,
  \href{http://dx.doi.org/10.1016/j.ultramic.2013.11.007}{{Optimized imaging
  using non-rigid registration}}, Ultramicroscopy 138 (2014) 46--56.
\newblock \href {http://dx.doi.org/10.1016/j.ultramic.2013.11.007}
  {\path{doi:10.1016/j.ultramic.2013.11.007}}.
\newline\urlprefix\url{http://dx.doi.org/10.1016/j.ultramic.2013.11.007}

\bibitem{Ophus2016}
C.~Ophus, J.~Ciston, C.~T. Nelson,
  \href{http://dx.doi.org/10.1016/j.ultramic.2015.12.002}{{Correcting nonlinear
  drift distortion of scanning probe and scanning transmission electron
  microscopies from image pairs with orthogonal scan directions}},
  Ultramicroscopy 162 (2016) 1--9.
\newblock \href {http://dx.doi.org/10.1016/j.ultramic.2015.12.002}
  {\path{doi:10.1016/j.ultramic.2015.12.002}}.
\newline\urlprefix\url{http://dx.doi.org/10.1016/j.ultramic.2015.12.002}

\bibitem{ning2017}
S.~Ning, T.~Fujita, A.~Nie, Z.~Wang, X.~Xu, J.~Chen, M.~Chen, S.~Yao, T.-Y.
  Zhang,
  \href{http://www.sciencedirect.com/science/article/pii/S0304399116302030}{Scanning
  distortion correction in stem images}, Ultramicroscopy\href
  {http://dx.doi.org/https://doi.org/10.1016/j.ultramic.2017.09.003}
  {\path{doi:https://doi.org/10.1016/j.ultramic.2017.09.003}}.
\newline\urlprefix\url{http://www.sciencedirect.com/science/article/pii/S0304399116302030}

\bibitem{elbaggari2017}
I.~{El Baggari}, B.~H. {Savitzky}, A.~S. {Admasu}, J.~{Kim}, S.-W. {Cheong},
  R.~{Hovden}, L.~F. {Kourkoutis}, {Commensurate Stripes and Phase Coherence in
  Manganites Revealed with Cryogenic Scanning Transmission Electron
  Microscopy}, ArXiv e-prints\href {http://arxiv.org/abs/1708.08871}
  {\path{arXiv:1708.08871}}.

\bibitem{zachman2016}
M.~J. Zachman, E.~Asenath-Smith, L.~A. Estroff, L.~F. Kourkoutis, Site-specific
  preparation of intact solid–liquid interfaces by label-free in situ
  localization and cryo-focused ion beam lift-out, Microscopy and Microanalysis
  22~(6) (2016) 1338–1349.
\newblock \href {http://dx.doi.org/10.1017/S1431927616011892}
  {\path{doi:10.1017/S1431927616011892}}.

\bibitem{wolf2014cryo}
S.~G. Wolf, L.~Houben, M.~Elbaum, Cryo-scanning transmission electron
  tomography of vitrified cells, nAture methods 11~(4) (2014) 423--428.

\bibitem{spoth2017dose}
K.~A. Spoth, K.~X. Nguyen, D.~A. Muller, L.~F. Kourkoutis, Dose-efficient
  cryo-stem imaging of whole cells using the electron microscope pixel array
  detector, Microscopy and Microanalysis 23~(S1) (2017) 804--805.

\bibitem{elad2017detection}
N.~Elad, G.~Bellapadrona, L.~Houben, I.~Sagi, M.~Elbaum, Detection of isolated
  protein-bound metal ions by single-particle cryo-stem, Proceedings of the
  National Academy of Sciences (2017) 201708609.

\bibitem{kimoto2010}
K.~Kimoto, T.~Asaka, X.~Yu, T.~Nagai, Y.~Matsui, K.~Ishizuka, Local crystal
  structure analysis with several picometer precision using scanning
  transmission electron microscopy, Ultramicroscopy 110~(7) (2010) 778--782.

\bibitem{Eerenstein2006}
W.~Eerenstein, N.~D. Mathur, J.~F. Scott, {Multiferroic and magnetoelectric
  materials}, Nature 442~(August) (2006) 759--766.
\newblock \href {http://dx.doi.org/10.1038/nature05023}
  {\path{doi:10.1038/nature05023}}.

\bibitem{Gruner1988}
G.~Gr\"uner, \href{https://link.aps.org/doi/10.1103/RevModPhys.60.1129}{The
  dynamics of charge-density waves}, Rev. Mod. Phys. 60 (1988) 1129--1181.
\newblock \href {http://dx.doi.org/10.1103/RevModPhys.60.1129}
  {\path{doi:10.1103/RevModPhys.60.1129}}.
\newline\urlprefix\url{https://link.aps.org/doi/10.1103/RevModPhys.60.1129}

\bibitem{tinkham1996introduction}
M.~Tinkham, Introduction to superconductivity, Courier Corporation, 1996.

\bibitem{VanHeel1992}
M.~van Heel, M.~Schatz, E.~Orlova, {Correlation functions revisited},
  Ultramicroscopy 46.

\bibitem{Foroosh2002}
H.~F. Shekarforoush, J.~B. Zerubia, S.~Member, M.~Berthod, {Extension of Phase
  Correlation to Subpixel Registration}, IEEE transactions on image processing
  11~(3) (2002) 188--200.

\bibitem{Watanabe2002}
K.~Watanabe, Y.~Kotaka, N.~Nakanishi, T.~Yamazaki, I.~Hashimoto, {Deconvolution
  processing of HAADF STEM images}, Ultramicroscopy 92 (2002) 191--199.

\bibitem{Nakanishi2002}
N.~Nakanishi, T.~Yamazaki, A.~Re, C.~Miran, K.~Watanabe, M.~Shiojiri,
  {Retrieval process of high-resolution HAADF-STEM images}, Japanese Society of
  Microscopy 51~(6) (2002) 383--390.

\bibitem{Recnik2005}
A.~Recnik, G.~Mobus, S.~Sturm, {IMAGE-WARP : A real-space restoration method
  for high-resolution STEM images using quantitative HRTEM analysis},
  Ultramicroscopy 103 (2005) 285--301.
\newblock \href {http://dx.doi.org/10.1016/j.ultramic.2005.01.003}
  {\path{doi:10.1016/j.ultramic.2005.01.003}}.

\bibitem{Sanchez2006}
A.~Sanchez, P.~Galindo, S.~Krets, M.~Falke, R.~Beanland, P.~Goodhew, {An
  approach to the systematic distortion correction in aberration-corrected
  HAADF images}, Journal of Microscopy 221~(1) (2006) 1--7.

\bibitem{Hytch1998}
M.~Hytch, E.~Snoeck, R.~Kilaas, {Quantitative measurement of displacement and
  strain fields from HREM micrographs}, Ultramicroscopy 74 (1998) 131--146.

\bibitem{Braidy2012}
N.~Braidy, Y.~Le, S.~Lazar, C.~Ricolleau,
  \href{http://dx.doi.org/10.1016/j.ultramic.2012.04.001}{{Correcting scanning
  instabilities from images of periodic structures}}, Ultramicroscopy 118
  (2012) 67--76.
\newblock \href {http://dx.doi.org/10.1016/j.ultramic.2012.04.001}
  {\path{doi:10.1016/j.ultramic.2012.04.001}}.
\newline\urlprefix\url{http://dx.doi.org/10.1016/j.ultramic.2012.04.001}

\bibitem{Jones2013}
L.~Jones, P.~D. Nellist, {Identifying and Correcting Scan Noise and Drift in
  the Scanning Transmission Electron Microscope}, Microscopy Microanalysis 19
  (2013) 1050--1060.

\bibitem{Zitova2003}
B.~Zitova, J.~Flusser, {Image registration methods : a survey}, Image and
  Vision Computing 21 (2003) 977--1000.
\newblock \href {http://dx.doi.org/10.1016/S0262-8856(03)00137-9}
  {\path{doi:10.1016/S0262-8856(03)00137-9}}.

\bibitem{modersitzki2004}
J.~Modersitzki, {Numerical methods for image registration}, Oxford University
  Press on Demand, 2004.

\bibitem{Robinson2004}
D.~Robinson, S.~Member, P.~Milanfar, S.~Member, {in Image Registration}, IEEE
  TRANSACTIONS ON IMAGE PROCESSING 13~(9) (2004) 1185--1199.

\bibitem{Aguerrebere2016}
C.~Aguerrebere, M.~Delbracio, A.~Bartesaghi, G.~Sapiro, {Fundamental Limits in
  Multi-Image Alignment}, IEEE TRANSACTIONS ON SIGNAL PROCESSING 64~(21) (2016)
  5707--5722.

\bibitem{vural2013}
E.~Vural, P.~Frossard, {Analysis of Descent-Based Image Registration ∗} 6~(4)
  (2013) 2310--2349.

\bibitem{Joyeux2002}
L.~Joyeux, P.~A. Penczek, {Efficiency of 2D alignment methods}, Ultramicroscopy
  92 (2002) 33--46.

\bibitem{Shatsky2009}
M.~Shatsky, R.~J. Hall, S.~E. Brenner, R.~M. Glaeser, {A Method for the
  Alignment of Heterogeneous Macromolecules from Electron Microscopy}, J.
  Struct. Biol. 166~(1) (2009) 67--78.
\newblock \href {http://dx.doi.org/10.1016/j.jsb.2008.12.008.A}
  {\path{doi:10.1016/j.jsb.2008.12.008.A}}.

\bibitem{Rubinstein2015}
J.~L. Rubinstein, M.~A. Brubaker, {Alignment of cryo-EM movies of individual
  particles by optimization of image translations}, Journal of Structural
  Biology 192 (2015) 188--195.
\newblock \href {http://dx.doi.org/10.1016/j.jsb.2015.08.007}
  {\path{doi:10.1016/j.jsb.2015.08.007}}.

\bibitem{Yankovich2014}
A.~B. Yankovich, B.~Berkels, W.~Dahmen, P.~Binev, S.~I. Sanchez, S.~A. Bradley,
  A.~Li, I.~Szlufarska, P.~M. Voyles, {Picometre-precision analysis of scanning
  transmission electron microscopy images of platinum nanocatalysts}, Nature
  Communications 5:4155 (2014) 1--7.
\newblock \href {http://dx.doi.org/10.1038/ncomms5155}
  {\path{doi:10.1038/ncomms5155}}.

\bibitem{Frank1981}
J.~Frank, A.~Verschoor, M.~Boublik, {Computer Averaging of Electron Micrographs
  of 40S Ribosomal Subunits}, Science 214 (1981) 1353.

\bibitem{allen1989}
M.~P. Allen, D.~J. Tildesley, {Computer simulation of liquids}, Oxford
  university press, 1989.

\bibitem{savitzky2017}
B.~H. {Savitzky}, I.~{El Baggari}, A.~S. {Admasu}, J.~{Kim}, S.-W. {Cheong},
  R.~{Hovden}, L.~F. {Kourkoutis}, {Bending and Breaking of Stripes in a
  Charge-Ordered Manganite}, ArXiv e-prints\href
  {http://arxiv.org/abs/1707.00221} {\path{arXiv:1707.00221}}.

\bibitem{Hovden2016}
R.~Hovden, A.~W. Tsen, P.~Liu, B.~H. Savitzky, I.~{El Baggari}, Y.~Liu, W.~Lu,
  Y.~Sun, P.~Kim, A.~N. Pasupathy, L.~F. Kourkoutis,
  \href{http://www.pnas.org/lookup/doi/10.1073/pnas.1606044113}{{Atomic lattice
  disorder in charge-density-wave phases of exfoliated dichalcogenides (1T-TaS
  2 )}}, Proceedings of the National Academy of Sciences 113~(41) (2016)
  11420--11424.
\newblock \href {http://dx.doi.org/10.1073/pnas.1606044113}
  {\path{doi:10.1073/pnas.1606044113}}.
\newline\urlprefix\url{http://www.pnas.org/lookup/doi/10.1073/pnas.1606044113}

\bibitem{Sheckelton2017}
J.~P. Sheckelton, K.~W. Plumb, B.~A. Trump, C.~L. Broholm, T.~M. McQueen,
  \href{http://dx.doi.org/10.1039/C6QI00470A}{Rearrangement of van der waals
  stacking and formation of a singlet state at t = 90 k in a cluster magnet},
  Inorg. Chem. Front. 4 (2017) 481--490.
\newblock \href {http://dx.doi.org/10.1039/C6QI00470A}
  {\path{doi:10.1039/C6QI00470A}}.
\newline\urlprefix\url{http://dx.doi.org/10.1039/C6QI00470A}

\bibitem{alvarez1999scale}
L.~Alvarez, J.~S{\'a}nchez, J.~Weickert, A scale-space approach to nonlocal
  optical flow calculations, in: International conference on scale-space
  theories in computer vision, Springer, 1999, pp. 235--246.

\bibitem{mundy2016}
J.~A. Mundy, C.~M. Brooks, M.~E. Holtz, J.~A. Moyer, H.~Das, A.~F. R{\'e}bola,
  J.~T. Heron, J.~D. Clarkson, S.~M. Disseler, Z.~Liu, et~al., Atomically
  engineered ferroic layers yield a room-temperature magnetoelectric
  multiferroic, Nature 537~(7621) (2016) 523--527.

\end{thebibliography}





\end{document}